\documentstyle[12pt]{article}
\input{psfig.tex}

\begin{document}
\begin{titlepage}

\hskip 12cm
\rightline{\vbox{\hbox{DFPD 98/TH 50}\hbox{CS-TH 5/98}
\hbox{Dec. 1998}}}
\vskip 0.6cm
\centerline{\bf HADRON DIFFRACTION DISSOCIATION AND}
\centerline{\bf THE TRIPLE POMERON VERTEX}
\vskip 1.0cm
\centerline{  R. Fiore$^{a\dagger}$, L. L. Jenkovszky$^{b\ddagger}$, F.
Paccanoni$^{c\ast}$}
\vskip .5cm
\centerline{$^{a}$ \sl  Dipartimento di Fisica, Universit\`a della Calabria
,}
\centerline{\sl Istituto Nazionale di Fisica Nucleare, Gruppo collegato di
Cosenza}
\centerline{\sl Arcavacata di Rende, I-87030 Cosenza, Italy}
\vskip .5cm
\centerline{$^{b}$ \sl  Bogoliubov Institute for Theoretical Physics,}
\centerline{\sl Academy of Sciences of the Ukrain}
\centerline{\sl 252143 Kiev, Ukrain}
\vskip .5cm
\centerline{$^{c}$ \sl  Dipartimento di Fisica, Universit\`a di Padova,}
\centerline{\sl Istituto Nazionale di Fisica Nucleare, Sezione di Padova}
\centerline{\sl via F. Marzolo 8, I-35131 Padova, Italy}
\vskip 1cm
\begin{abstract}
Hadron diffraction dissociation is considered in the dipole 
Pomeron model where the Pomeron is represented by a double 
pole in the $J$-plane. We find that unitarity is satisfied 
without decoupling of the triple Pomeron vertex. The reaction
$\bar{p}+p \to \bar{p}+X$ is analysed.
\end{abstract}
\vskip .5cm
\hrule
\vskip .3cm
\noindent

\noindent
$^{\diamond}${\it Work supported by the Ministero italiano
dell'Universit\`a e della Ricerca Scientifica e Tecnologica and by the 
INTAS}
\vfill
$\begin{array}{ll}
^{\dagger}\mbox{{\it email address:}} &
   \mbox{FIORE~@CS.INFN.IT}
\end{array}
$

$ \begin{array}{ll}
^{\ddagger}\mbox{{\it email address:}} &
   \mbox{JENK~@GLUK.APC.ORG}
\end{array} 
$

$ \begin{array}{ll}
^{\ast}\mbox{{\it email address:}} &
   \mbox{PACCANONI~@PADOVA.INFN.IT}
\end{array}
$
\vfill
\end{titlepage}
\eject
\textheight 210mm
\topmargin 2mm
\baselineskip=24pt

{\bf 1. INTRODUCTION}

\vskip 1.5 cm
                                                                            
The inclusive process of hadron diffraction dissociation has been
discussed extensively in the literature 
\cite{bib1,bib2,bib3,bib4,bib5,bib6}.
In this paper we will be mainly concerned with the violation of
unitarity appearing in the Regge theory, where the diffraction
cross-section rises faster than the total one.

Solutions to this problem have been proposed based on different
unitarization recipes. Eikonal corrections \cite{bib7}
succeded, for example, in reproducing the main features of
single diffraction at high energy obtaining a weak energy
dependence of the relevant cross-sections.
The inclusion of cuts in the Regge theory \cite{bib8} has the
same effect while a different approach, based on the 
unitarization of the Pomeron flux, has been advocated by
\cite{bib9} and \cite{bib10}. Alternative ways of unitarization
have been considered in \cite{bib11}.
All the above approaches are
based on a supercritical Pomeron input.

A different solution is provided by successful models that
are based on the following assumption \cite{bib12}.
Suppose that asymptotically the absorptive part in the $s$-channel,
A(s,t), goes like
\begin{displaymath}
A(s,t) \propto \beta_1(t)\beta_2(t) s^{\alpha(t)}
[h(t) \ln s+ C]~,
\end{displaymath}
then the partial wave amplitude presents a simple and a double
pole in the complex J-plane.
The amplitude for the Pomeron exchange can then be written as
\begin{equation}
T(s,t) \propto -\frac{(-is)^{\alpha(t)}}{\sin(\pi\alpha(t)/2)}
\beta_1(t)\beta_2(t)[h(t)(\ln s-i\frac{\pi}{2})+C]~,
\label{z1}
\end{equation}
where constant terms have been collected in C. 

Eq.~(\ref{z1}) derives
from an ansatz on the form of the Regge residues and different
expressions, as far as the $t$-dependence of h(t) is concerned,
can be found in the literature. As an example, in a dual model,
if the residue of the simple pole has the form $\beta(\alpha(t))$,
the residue of the double pole will be given by
$\int\beta(\alpha(t))\,d\alpha+ const$. This formalism gives  
an excellent description of $p-p$ and $p-\bar{p}$ elastic 
scattering including the dip \cite{bib12}. We will show that, starting from
Eq~.(\ref{z1}), it is possible to satisfy unitarity at the Born level,
without eikonalization.

In the following Sections we apply this model to single
diffractive dissociation following, and developing further,
the previous work \cite{bib13} on this subject.

{\bf 2. DIFFRACTIVE DISSOCIATION}

Consider first the process $a+b\to c+X$ with the exchange of 
Regge trajectories ${i}$. From the Mueller discontinuity formula 
\cite{bib14} we get
\begin{equation}
\pi E_c\frac{d^3\sigma}{d\vec{p}_c}=
\frac{1}{16\pi s}\sum_X\left|\sum_i\beta^i_{a\bar{c}}(t)
\xi_i(t)F^{ib\to X}(M^2,t)\left(\frac{s}{M^2}\right)^{\alpha_i
(t)}\right|^2   
\label{z2}
\end{equation}
in the usual Regge pole model. $M^2$ is the squared mass of the
unrevealed state $X$ and $\alpha_i(t)$ represents the Regge trajectory
exchanged. In the following $i=P, f,\pi$, while the $\omega$ trajectory will 
be neglected on the basis of historical fits \cite{bib1,bib15}. $P$ stands for 
the Pomeron trajectory and 
$\xi_i(t)=(1\pm\exp(-i\pi\alpha_i(t))/\sin(\pi\alpha_i(t))$ is the signature.
\vskip 0.3cm
In the dipole Pomeron approach, Eq.~(\ref{z2}) becomes
\begin{eqnarray}
& &  \pi E_c\frac{d^3\sigma}{d\vec{p}_c}=
\nonumber \\
& &  \frac{1}{16\pi s}\sum_X\left| \beta^{P}_{a\bar{c}}(t)
\left(-i\frac{s}{M^2}\right)^{\alpha_{P}(t)}\left[h(t)(\ln
\frac{s}{M^2}-i\pi/2)+C\right]F^{P b\to X}(M^2,t) \right.+
\nonumber \\
& &\left. \sum_{i\neq P} \beta^i_{a\bar{c}}\xi_i(t) F^{ib\to X}(M^2,t)
\left(\frac{s}{M^2}\right)^{\alpha_i(t)} \right|^2.
\label{z3}
\end{eqnarray}

Since $\pi$ contributes in a different kinematical region with respect to 
$P$ and $f$, interference terms between $\pi$ and $P$,
$f$ are suppressed. Hence in Eq.~(\ref{z3}) the sum over $i$ refers only
to $f$, and the $\pi$ contribution will be chosen as in
\cite{bib6,bib15,bib16,bib18}:
\begin{equation}
\frac{1}{4\pi}\frac{g^2_{\pi ac}}{4\pi}\frac{(-t)}{(t-\mu^2)^2}
\left(\frac{s}{M^2}\right)^{2\alpha_{\pi}(t)-1}G^2(t)
\sigma^{\pi p}_T(M^2)~,
\label{z4}
\end{equation}
where
\begin{displaymath}
G(t)=\frac{2.3-\mu^2}{2.3-t}~.
\end{displaymath}
Since $a=c=\bar{p}$ and $b=p$, we have also 
\begin{displaymath}
\frac{g^2_{\pi pp}}{4\pi}=14.6~.
\end{displaymath}

The $f$-exchange can be treated in the approximation suggested in 
\cite{bib19,bib20} with the result that the $s$-dependence
will be modified in Eq.~(\ref{z3}) as
\begin{equation}
\left(\frac{s}{M^2}\right)^{2\alpha_{P}(t)}[(h(t)\ln\frac{s}{M^2}
+C)^2+\frac{\pi^2}{4}h^2(t)+R(s,t)]~,     
\label{z5}
\end{equation}
where
\begin{eqnarray}
R(s,t) &=& k\left([h(t)\ln\frac{s}{M^2}+C] \cos\left(\frac{\pi a(t)}{2}
\right)- \right.
\nonumber \\
 & &  \left. \frac{\pi h(t)}{2}\sin\left(\frac{\pi 
a(t)}{2}\right)\right) \left(\frac{s}{M^2}
\right)^{-a(t)}+k^2\left(\frac{s}{M^2}\right)^{-2a(t)}
\label{z6}
\end{eqnarray}
and  $a(t)$ is the difference between the
$P$ and $f$ trajectories:
\begin{displaymath}
a(t)=\alpha_{P}(t)-\alpha_f(t)= a(0)-\delta t.
\end{displaymath}
Typical values are $a(0)\simeq 0.5$ and $\delta\simeq 0.65$.
In \cite{bib19} it is quoted for $k$ a value near $7.8$.

{\bf 3. THE TRIPLE POMERON}

Let us consider now the triple Pomeron contribution to Eq.~(\ref{z3}),
with $\alpha_{P}(t)=1+\alpha't$ and $\alpha'=0.25\,GeV^{-2}$,
\begin{eqnarray}
& &\frac{1}{16\pi s}[\beta^{P}_{a\bar{c}}(t)]^2\left(\frac{s}{M^2}
\right)^{2\alpha_{P}(t)}\times 
\nonumber \\
 & &  \left[(h(t)\ln\frac{s}{M^2}+C)^2+\frac{\pi^2}{4}h^2(t)\right]
Im\,T^{P b}(M^2,t,\alpha_{P}(t),t_{b\bar{b}}=0)~,
\label{z7}
\end{eqnarray}
where, according to the dipole Pomeron model,
\begin{equation}
Im\,T^{P b}=\sigma_0\,(M^2)^{\alpha_{P}(0)}(\lambda+
h(0)\ln M^2) g(t)~,   
\label{z8}
\end{equation}
$g(t)$ being the triple Pomeron coupling and, for simplicity sake,
the same function $h(t)$ has been considered. A term, decreasing with
$M^2$, could well be present in (\ref{z8}) if we consider also secondary
trajectories in $P-b$ scattering. Hence Eq.~(\ref{z8}) will be valid
only for $M^2$ sufficiently large.

The presence of the function $h(t)$ in the contribution (\ref{z7}) and 
Eq.~(\ref{z8}) is characteristic of the model considered. In terms of 
partial waves, if
\begin{equation}
\frac{d}{dJ}\left[\beta^{P}_{a\bar{c}}(J,t)F^{P b\to X}
(J,M^2,t)\right]_{J=\alpha_{P}(t)}  
\label{z9}
\end{equation}
is the coefficient of the simple pole, then the coefficient of
$\ln s$,
\begin{equation}
\left. \beta^{P}_{a\bar{c}}(J,t)F^{P b\to X}(J,M^2,t)
\right|_{J=\alpha_{P}(t)}~,
\label{z10}
\end{equation}
can be obtained from the expression (\ref{z9}) by integration provided a 
phenomenological form for the residue of the simple pole is available.
In absence of a reliable input we must resort to other
constraints, for example we can impose that unitarity is satisfied.

By integrating Eq.~(\ref{z3}) over $t$ and $M^2$ we get the single
diffractive cross-section, $\sigma_{SD}$. The constraint
$\sigma_{SD}<\sigma_T$ for all values of $s$ requires that 
$h(t)\propto (-t)^{\gamma}$. A lower bound for $\gamma$ will
be discussed in the next Section.
Anyway, explicit examples where $h(t)$ must vanish as $(-t)$
when $t$ goes to zero, can be readily found. Assuming the
simple form $h(t)=h(-t)^{\gamma}$,with $h$ constant, and taking the 
phenomenological expression $\exp(bt)$ for the Pomeron-proton
vertex, the triple Pomeron contribution (\ref{z7}) becomes
\begin{equation}
\frac{h^2\sigma_0M^2\lambda g(0)}{16\pi s}e^{2(bt+\alpha_{P}(t)
\ln(s/M^2))}
\left[((-t)^{\gamma}\ln\frac{s}{M^2}+C)^2+
\frac{\pi^2}{4}(-t)^{2\gamma}\right]~,
\label{z11}
\end{equation}
where the triple Pomeron vertex has been considered as
constant, according to experiments \cite{bib6,bib17}. $C/h$
has been renamed as $C$.

The final form of the differential cross-section is 
\begin{eqnarray}
\frac{d^2\sigma}{dt\,dM^2}  &=&  \frac{A}{M^2}
e^{2(b+\alpha'\ln(s/M^2))t}
\left[((-t)^{\gamma}\ln\frac{s}{M^2}+
C)^2+\frac{\pi^2}{4}(-t)^{2\gamma}+R(s,t)/h^2\right]+
\nonumber  \\
& &  \frac{1}{4\pi}\frac{g^2}{4\pi M^2}\frac{(-t)}{(t-\mu^2)^2}G^2(t)
\left(\frac{s}{M^2}\right)^{2\alpha_{\pi}(t)-2} \sigma^{\pi p}_T
(M^2)~,
\label{z12}
\end{eqnarray}
with
\begin{displaymath}
A=\frac{h^2\sigma_0\lambda g(0)}{16\pi}~.
\end{displaymath}
In $R(s,t)$ the substitution $h(t)=h\,(-t)^{\gamma}$ must be
performed and C has been redefined accordingly.
As far as the other parameters are concerned, $b$ will be fixed
from p-p elastic scattering, e.g. $b=2.25\,GeV^{-2}$, and 
$\sigma^{\pi p}_T(M^2)$ in the dipole Pomeron model can
be written as
\begin{equation}
\sigma^{\pi p}_T(M^2)=3.62+2.55\ln(M^2)+38.89 (M^2)^{-0.34}
\label{z13}
\end{equation}
inspired by the parametrization used in \cite{bib21}.
Moreover
\begin{displaymath}
\alpha_{\pi}(t)=0.9\,t~.
\end{displaymath}
Since the form of $h(t)$ is determined only near $t=0$, it is well
possible that the $t$-dependence of the cross-section should
be corrected. Hence a different value of $b$ could be required
from the experimental data.

{\bf 4. THE PARAMETER $\gamma$ AND ITS LOWER BOUND}

In order to make explicit the unitarity constraint on the single
diffraction cross-section we must integrate Eq.~(\ref{z12}) in the
variables $t$ and $M^2$. By integrating in the variable $t$ 
in the range $(-\infty, 0)$ we get the contribution to
$d\sigma/dM^2$ from the pion and $f$ trajectories given in
Appendix.

More important for the bound is the integral over the same 
$t$-interval of the Pomeron contribution \ref{z11}:
\begin{equation}
\frac{A}{M^2} \left[ \Gamma(2\gamma+1)\frac{(\ln s/M^2)^2+
\pi^2/4}{p^{2\gamma+1}}+ 
 2\Gamma(\gamma+1)C\frac{\ln s/M^2}{p^{\gamma+1}}+\frac{C^2}{p}
\right]~,
\label{z14}
\end{equation}
where
\begin{displaymath}
p=2\left(b+\alpha' \ln\frac{s}{M^2}\right)~.
\end{displaymath}
When integrated over $M^2$, between the limits $\zeta=M_{min}^2$ and
$\rho s$ \footnote{CDF \cite{bib22} chooses $\zeta=1.4\,GeV^2$ and
$\rho=0.15$.}, the expression (\ref{z14}) gives
\begin{eqnarray}
A\int^{\ln(s/\zeta)}_{\ln(1/\rho)}\,dx
\left[\Gamma(2\gamma+1)\frac{x^2+\pi^2/4}{(2(b+\alpha' x))^{2\gamma+1}}
\right. &+& 
\nonumber  \\
 2\Gamma(\gamma+1)C\frac{x}{(2(b+\alpha' x))^{\gamma+1}} &+& \left.
\frac{C^2}{2(b+\alpha' x)}\right]~,
\label{z15}
\end{eqnarray}
whose increase must be bounded by $\ln s$, that is the behaviour
in $s$ of the total cross-section in the dipole Pomeron model.
The smallest allowed value of $\gamma$ is  
$\gamma_{min.}\geq 1/2$.
This can be easily seen by noticing that
\begin{displaymath}
\int_l^u\,dx \frac{x^n}{(1+ax)^\nu}=
\frac{1}{a^{n+1}}\sum_{m=0}^n \left(\begin{array}{c} n \\ m
\end{array} \right) \left. \frac{1}{m-\nu+1}(1+ax)^{m-\nu+1}\right|^u_l
\end{displaymath}
has the asymptotic behaviour $u^{n-\nu+1}+O(u^{n-\nu})$, for $\nu$ not
integer, and that $\nu-n\geq 0$.

Hence, the parameter $\gamma$, in general, must satisfy: $\gamma\geq 1/2$.
This inequality is necessary to avoid terms, violating unitarity,
that rise faster than $\ln s$.
It is important to notice that the triple Pomeron
contribution does not vanish at $t=0$ because of the presence of the
constant $C$. 

{\bf 5. COMPARISON WITH DATA AND CONCLUSIONS}

When comparing the model with experimental data we find two kinds
of problems. The first one is related to the experimental definition
of single diffraction dissociation. The great variety of 
phenomenological models adopted from different experimental groups,
in order to extract the published data, makes difficult the test of any
new model. Moreover, integrated cross-section do not refer to
the same intervals of $M^2$ and $t$, for different experimental
analyses.

The second kind of problem resides in our parametrization and is
strongly related to the first one. While the pion contribution
can be fixed as in Section 2, the parameters relative to the $f$
trajectory must be refitted since the Pomeron contribution is
quite different from the one proposed in \cite{bib19,bib20}.
Moreover, the parameter $\gamma$ requires fine tuning, since its
value must be larger than $\frac{1}{2}$, and the choice of the
function $h(t)$ has been made only on the basis of its small
$t$ behaviour.

In view of these difficulties, we simplify the analysis by 
neglecting the $f$-contri\\bution
together with the $P-f$ interference term.  
The $\pi$ contribution has no free parameters and we are left
with three parameters for the Pomeron and a possible
correction to the slope $b$.

\begin{figure}
\centerline{\psfig{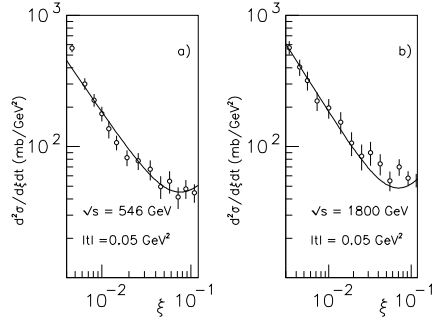}}
\caption[]{Differential cross sections $d^2\sigma/d\xi\,dt$
vs $\xi$, data are from \cite{bib22} compiled by \cite{bib17}. 
The solid curves represent the result of the model.}
\label{fig1}
\end{figure}

By comparing the slope of $d^2\sigma/d\xi dt$, at fixed
$\xi=M^2/s=0.035$ and in the interval $0.05<-t<0.15$, with the
CDF \cite{bib22} and UA8 \cite{bib18} best fits, we find that
at $\sqrt{s}=1800\,GeV$ the prediction of
the model is well within the experimental errors if we choose
$\gamma=0.95$ and keep the previous value for $b$,
$b=2.25\,GeV^{-2}$.  At $\sqrt{s}=
546\,GeV$ the slope agrees with \cite{bib18} and simulates the
effect of a non linear Pomeron trajectory \cite{bib18,bib24}.
Larger values of the parameter $\gamma$ are compatible with
the experimental slope and bring forth a flatter total
diffractive cross section, as can be seen from the Pomeron contribution 
(\ref{z15}).

Then we performed a fit, at fixed $|t|=0.05\,GeV^2$, of the
CDF data \cite{bib22}, taken from the providential compilation
in \cite{bib17}. Figs. \ref{fig1}a and 1b show that the proposed
model is adequate as far as the $\xi$ dependence is
concerned. Only the data above the resonance region, $M^2>5\,GeV^2$,
and in the coherence region \cite{bib6}, $M^2/s<0.1$, have
been considered. The parameters of the fit are $A=0.1098 and 
\,C=4.557$, with strong correlations between them.

\begin{figure}
\centerline{\psfig{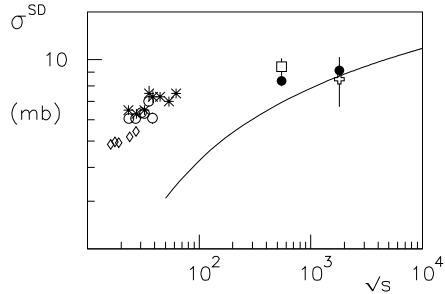}}
\caption[]{Total single diffraction cross section vs
$\sqrt{s}$ compared with the prediction of the model. Data are 
taken from the compilation of \cite{bib17}.}
\label{fig2}
\end{figure}

If we consider the data at smaller $s$, $\sqrt{s}=20\,GeV$,
\cite{bib23} compiled in \cite{bib17}, the calculated cross section 
is well below the experimental points. We attribute such a 
discrepancy to the neglect of the
$f$ trajectory that, in this energy region, plays an important
role in $p-\bar{p}$ interaction. The $f$ contribution could also
modify sensibly the imaginary part of the $P-b$ amplitude in
Eq.~(\ref{z3}).

In Fig. \ref{fig2} the total single diffractive cross section $\sigma_{sd}$,
for the process $p(\bar{p})+p\to p(\bar{p})+X$, is shown as a
function of $\sqrt{s}$. 
We have considered the experimental
data of \cite{bib22,bib25,bib26,bib27,bib28} from the compilation given in 
\cite{bib17} where some data have been corrected in order to obtain
the diffraction cross section for $\xi\leq 0.05$. 
 Again, the absence of the $f$
trajectory makes the cross section too small at low energy.
Analogous results have been obtained in  a theoretical model
\cite{bib7} and can be deduced from a phenomenological
parametrization \cite{bib22}.
We notice that the cross-section rises as $(\ln s)^{0.1}$ in this model.
A more complete fit of all the data will be considered
elsewhere.

From the theoretical pointof view, the result in Eq.~(12) possesses two
important properties that seem required by the data \cite{bib17}.
First, exact factorization, typical of the Regge pole model, is
lost in the dipole Pomeron approach. Second, for $t=0$, the
Pomeron and pion contributions are independent of $s$.
Finally we remark that this model respects the unitarity condition
without decoupling of the triple Pomeron vertex. The total
diffractive cross section presents a slower rise than the total
$p-\bar{p}$ cross section that, in turn, satisfies the
Froissart bound. 

\section{Appendix}

By evaluating the integral over $t$ of the expression (\ref{z4}) in the range
$(-\infty,0)$ we get the contribution of the $\pi$-trajectory
to $d\sigma/dM^2$:
\begin{eqnarray*}
\frac{1}{4\pi}\left(\frac{g^2}{4\pi}\right)\frac{1}{M^2}
\left(\frac{s}{M^2}\right)^{-2}\sigma^{\pi p}_T(M^2) 
 \left[e^{\mu^2p_{\pi}}E_1(\mu^2p_{\pi})\left(\frac{2.3+\mu^2}{2.3
-\mu^2}+\mu^2p_{\pi}\right) \right. &-&         \\ 
\left. e^{2.3 p_{\pi}}E_1(2.3 p_{\pi})
\left(\frac{2.3+\mu^2}{2.3-\mu^2}-2.3 p_{\pi}\right)-2
\right]~,
\end{eqnarray*}
where $p_{\pi}=2\alpha_{\pi}(t)\ln(s/M^2)/t$.
$E_1(x)=-E_i(-x)$ is the exponential integral \cite{bib29}. 

From the $f$-trajectory the contribution is
\begin{eqnarray*}
& &   \frac{kA}{hM^2}\left[\left(
\frac{s}{M^2}\right)^{-a(0)}\left(\frac{\Gamma(\gamma+
1)}{(p_1^2+(\pi\delta/2)^2)^{(\gamma+1)/2}}V(s/M^2)+ \right. \right. \\
& & \left. \left.  \frac{C}{\sqrt{p_1^2+(\pi\delta/2)^2}}W(s/M^2)\right) +
 \left(\frac{s}{M^2}\right)^{-2 a(0)}\frac{k}{hp_2}\right]~,
\end{eqnarray*}
where
\begin{eqnarray*}
V(s/M^2)  &=&  \cos\left[\frac{\pi a(0)}{2}+ (\gamma+1)\tan^{-1}
\left(\frac{\pi\delta}{2p_1}\right)\right]\left(\ln\frac{s}{M^2}\right)-
 \\
& &  \frac{\pi}{2} \sin\left[\frac{\pi a(0)}{2}+ (\gamma+1)\tan^{-1}\left(
\frac{\pi\delta}{2p_1}\right)\right],
\end{eqnarray*}

\begin{displaymath}
W(s/M^2)=\cos\left[\frac{\pi a(0)}{2}+\tan^{-1}\left(\frac{\pi
\delta}{2 p_1}\right)\right],
\end{displaymath}
and
\begin{displaymath}
p_1=2b+(2\alpha'+\delta)\ln\frac{s}{M^2} \mbox{      },~~~~~~
p_2=2\left(b+(\alpha'+\delta)\ln\frac{s}{M^2}\right).
\end{displaymath}
 
\vskip 1.5cm
\underline{Acknowledgements}:
We thank K. Goulianos for useful correspondence on the subject of
this work. One of us (L.L.J.) is grateful to the Dipartimento
di Fisica della Universit\`a della Calabria and to the
Istituto Nazionale di Fisica Nucleare-Sezione di Padova e
Gruppo collegato di Cosenza for their warm hospitality and
financial support.

\vfill\eject
\end{document}